\documentclass[structabstract]{aa}  
\usepackage{graphicx}
\usepackage{txfonts}
\usepackage{natbib}
\bibpunct{(}{)}{;}{a}{}{,} 

\newcommand{\prp}    {${\rlap.}^{\prime}$}
\newcommand{\grp}    {${\rlap.}^{\circ}$}
\newcommand{\pri}    {${\rlap.}^{\prime \prime}$}
\newcommand{\rl}     {${\rlap.}^{s}$}

\newcommand{\ltsima} {$\; \buildrel < \over \sim \;$}
\newcommand{\simlt}  {\lower.5ex\hbox{\ltsima}}            
\newcommand{\gtsima} {$\; \buildrel > \over \sim \;$}
\newcommand{\simgt}  {\lower.5ex\hbox{\gtsima}}            

\begin{document}
%
\title{Counterpart candidates to the unidentified Fermi source 0FGL~J1848.6$-$0138}


   \author{P. L. Luque-Escamilla\inst{1}
          \and
          J. Mart\'{\i}\inst{2}
          \and
          A. J.  Mu\~noz-Arjonilla\inst{2}
          \and
          J. R. S\'anchez-Sutil\inst{2}
          \and
          J. A. Combi\inst{2,3}
          \and
          E. S\'anchez-Ayaso\inst{2}
          }

   \institute{Dpto. de Ing. Mec\'anica y Minera, EPSJ,
Universidad de Ja\'en, Campus Las Lagunillas s/n, Edif. A3, 23071 Ja\'en, Spain\\
              \email{peter@ujaen.es}
         \and
Departamento de F\'{\i}sica, EPSJ,
Universidad de Ja\'en, Campus Las Lagunillas s/n, Edif. A3, 23071 Ja\'en, Spain\\
             \email{jmarti@ujaen.es, ajmunoz@ujaen.es, jrssutil@ujaen.es, jcombi@ujaen.es, esayaso@ujaen.es}
         \and
Facultad de Ciencias Astron\'omicas y Geof\'{\i}sicas, Universidad
Nacional de La Plata, Paseo del Bosque, B1900FWA La Plata, Argentina\\
             }

   \date{Received ; accepted }

 
  \abstract
{}
   {We aim here to contribute to the identification of unassociated
bright sources of gamma-rays in the recently released catalogue obtained
by the Fermi collaboration.}
   {Our work is based on a extensive cross-identification of sources from different
wavelength catalogues and databases.}
   {As a first result, 
we report the finding of a few counterpart candidates inside
the 95\% confidence error box of the Fermi LAT unidentified gamma-ray 
source 0FGL J1848.6$-$0138. The globular cluster GLIMPSE-C01 remarkably stands out among 
the most peculiar objects consistent with the position
uncertainty of the gamma-ray source and with a conceivable physical scenario for gamma-ray
production. The Fermi observed spectrum is compared against theoretical predictions in the literature
making the association plausible but not yet certain due to its low X-ray to gamma-ray luminosity ratio.
Other competing counterparts are also discussed.
In particular, we pay a special attention to
a possible Pulsar Wind Nebula inside the Fermi error box whose nature is yet to be confirmed.
}
   {Both a globular cluster and an infrared source resembling a Pulsar Wind Nebula have been found
in positional agreement with 0FGL J1848.6$-$0138.
In addition, other interesting objects in the field are also reported.
Future gamma-ray observations will narrow the position uncertainty and we hope to eventually confirm
one of the counterpart candidates reported here. If GLIMPSE-C01 is confirmed,
together with the possible Fermi detection of the well known globular cluster 47 Tuc, 
then it would provide strong support to theoretical predictions of
globular clusters as gamma-ray sources.
}


\titlerunning{Counterpart candidates to 0FGL J1848.6$-$0138}
   \maketitle
%

\section{Introduction}

The collaboration operating the Fermi Large Area Telescope (LAT) has recently
released a first catalogue of highly-significant gamma-ray sources based on the first three months
of observation  \citep{fermicat}. The LAT instrument on board Fermi is
extensively described in Atwood et al. (2009) and references therein. Its performance represents a significant
step forward with respect to previous gamma-ray space missions, such as the COMPTON-GRO satellite, whose
poor angular resolution rendered very difficult the identification of most sources.
Among the 205 Fermi bright sources so far reported with significance of 10-$\sigma$ or higher, 38 of them remain unassociated
with any known object at lower energies. 

We have carried out a cross-identification search of these
unidentified Fermi sources with different catalogues and databases. The typical 95\% confidence error radius
of bright Fermi sources is within 10 to 20 arc-minute. Despite the remarkable improvement
as compared to past missions, it is not unusual to find several 
counterpart candidates
consistent with Fermi error circles.
However, in a few occasions we do find
one or a few potentially interesting objects which could be responsible for the gamma-ray detection.
One of these cases corresponds to the Fermi source
\object{0FGL J1848.6$-$0138}, whose error box contains  
the globular cluster \object{GLIMPSE-C01} \citep{kob05} among other possible counterparts.

In this {\it Letter}, 
we first devote our attention to the evidence in support of a globular
cluster (GC) association both from the observational and theoretical point of view.
The possibility of GCs as a new class of gamma-ray sources was predicted
many years ago by different authors \citep{chen91, tav93}. The production of gamma-ray photons is expected
to be powered by a population of millisecond radio pulsars (MSPs) inside the GC, estimated to be of $\sim10$-$10^2$
order. These pulsars continuously inject relativistic leptons into the GC medium either from their
inner magnetospheres or accelerated in the shock waves created by the collision of individual pulsar winds.
Recent theoretical predictions to assess the chances of detection by the new generation of Cherenkov and
satellite gamma-ray telescopes assume that gamma-ray emission is produced by inverse
Compton scattering of these leptons with the stellar and microwave background radiation \citep{bs07}.
The feasibility of this physical scenario is further enhanced by the suggested identification
of the well known GC \object{NGC~104} (47 Tuc) with one of the Fermi gamma-ray sources, i.e.,
\object{0FGL J0025.1$-$7202} \citep{fermicat}.

\begin{figure*}
\includegraphics[angle=0, scale=0.55]{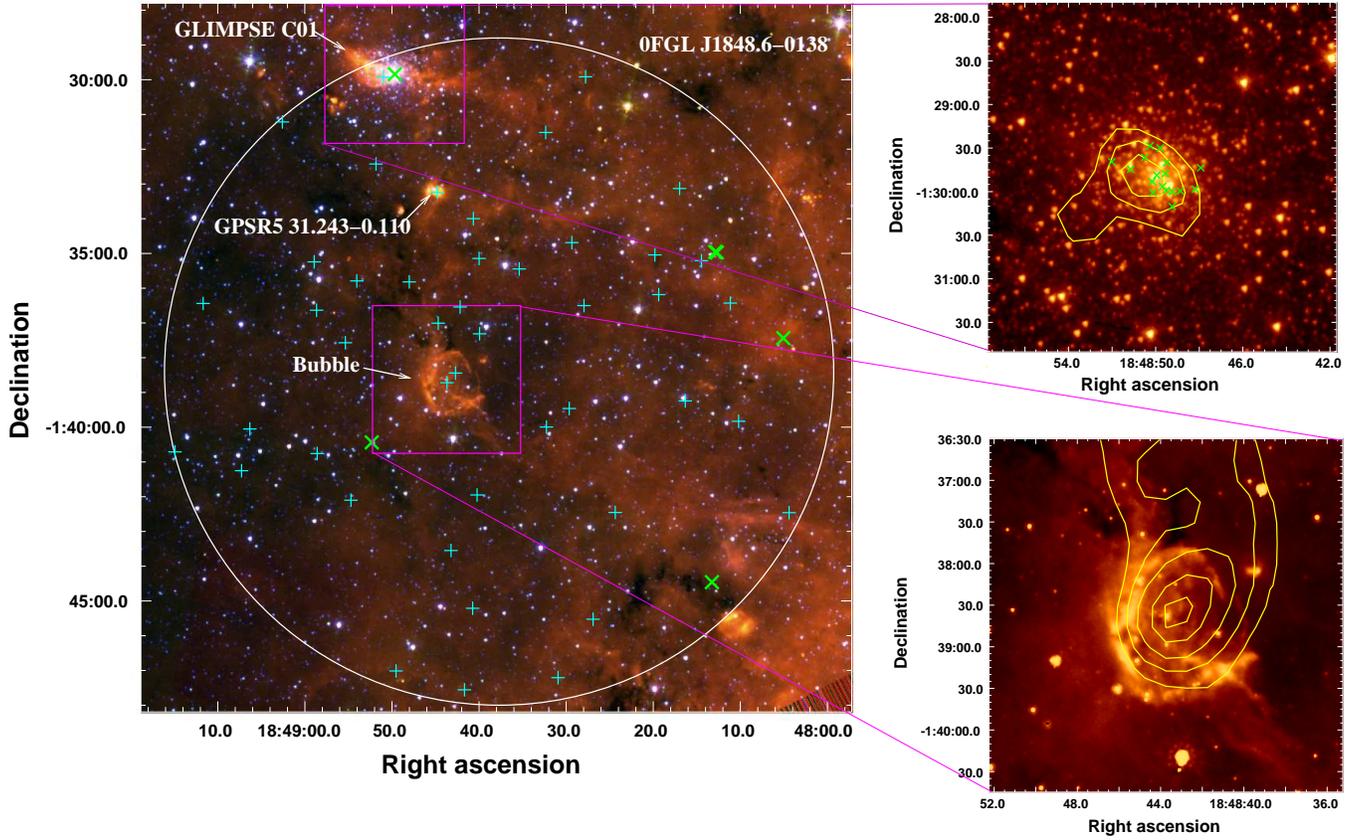}
\caption{{\bf Left.} Tri-colour GLIMPSE image covering the 95\% confidence position
of the gamma-ray source 0FGL J1848.6$-$0138 shown as a white circle.
Blue crosses represent radio sources in the field from the NVSS catalogue and
green crosses mark the location of X-ray sources detected by {\it XMM-Newton}.
{\bf Right.} The right panels illustrate an enlarged view for both the GC (3.6 $\mu$m, top) 
and the bubble-like object (8 $\mu$m, bottom),
including their respective NVSS radio emission as yellow contours with
angular resolution of $45^{\prime\prime}$. The emission levels shown correspond
to 3, 4 and 5 times the local rms noise of 1 mJy for the GC and 3, 9, 18, 30 and 40 times for the apparent bubble source.
Small green crosses are {\it Chandra} X-ray sources.
GLIMPSE-C01 appears as a faint radio source and contains numerous X-ray sources detected by {\it Chandra} marked as
small green crosses. On the other hand, the proposed
bubble is a strong radio emitter and its possible nature is discussed in the text.
}
\label{cluster_image}
\end{figure*}

Secondly, we also report other alternative counterpart candidates
inside the \object{0FGL J0025.1$-$7202} error circle but whose nature is not yet fully established.
It is interesting that one of them could be a Pulsar Wind Nebula (PWN). The association of gamma-ray
sources with these late products of stellar evolution is a well established fact and the Crab nebula
is the most prototypical example. Whether a PWN or a less conventional kind of counterpart, such
a GC, is behind \object{0FGL J0025.1$-$7202} is an issue yet to be solved.

\section{Cross-identification of Fermi and multi-wavelength archival data}

We initially performed a quick cross-iden\-tification of unassociated Fermi sources
with different radio, infrared and X-ray catalogues and databases, such as the
NRAO Very Large Sky Survey \citep{con98}, hereafter NVSS, the Spitzer/IRAC GLIMPSE Survey \cite{b03} and
the XMM-Newton Serendipitous Source Catalog, 2nd Version,
2XMM\footnote{http://heasarc.gsfc.nasa.gov/FTP/xmm/data/catalogues/\\2XMMcat−v1.0.fits.gz}, respectively.

As a result, the case of \object{0FGL J1848.6$-$0138} stands out due to the obvious
presence of the GC \object{GLIMPSE-C01} ($l=$31\grp 3, $b=-$0\grp 1) inside its Fermi 9\prp 6 radius of 95\% confidence.
In left panel of Fig. \ref{cluster_image} we show the composite (3.6, 5.8 and 8.0 $\mu$m bands) GLIMPSE image of the field where the GC
is clearly detected. Moreover, it appears as a faint source and contains numerous X-ray emitters detected by {\it Chandra} (see top right panel of same Fig. 
\ref{cluster_image}).

Encouraged by this finding, a closer inspection of GLIMPSE data revealed
other potentially interesting sources
consistent with the \object{0FGL J1848.6$-$0138} position.
Among them there is an almost circular bubble, or shell-like object,
located at $RA=18^h 48^m 43^s$ and $DEC=-01^{\circ}$ 38\prp 7 and being a very strong radio source. The bottom right panel
of Fig. \ref{cluster_image} shows and enlarged view of it.  Its morphology is reminiscent
of a PWN, but we cannot confidently classify it yet as evidenced in the following discussion.
The ultracompact HII region GPSR5 31.243$-$0.110 is also consistent with the positon of the Fermi source.

\section{Discussion}

This section is devoted to assess all the different counterpart alternatives reported in this paper.

\subsection{The GC GLIMPSE-C01 as a candidate counterpart}

\begin{figure}
\includegraphics[angle=0, scale=0.17]{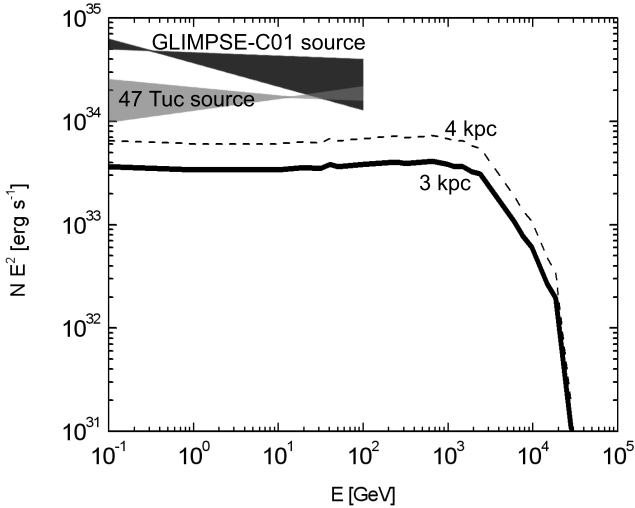}
\caption{Comparison of the observed Fermi emission for 0FGL J1848.6$-$0138
and 0FGL J0025.1$-$7202 in the GLIMPSE C-01 and 47 Tuc globular cluster fields, respectively,
with some of the gamma-ray predictions discussed in the text \cite{bs07}.
The shaded regions correspond to the spectral fit uncertainty and
reasonable distances to both clusters of 3 and 4 kpc are assumed.}
 \label{comparison}
\end{figure}

This heavily obscured ($A_V \simeq 15 \pm 3$)
cluster was originally reported and studied in detail a few years ago by Kobulnicky et al. (2005).
It appears to have an estimated mass of at least $\sim 10^5$ $M_{\odot}$ and an age of a few gigayears.
The distance to \object{GLIMPSE-C01} is still highly uncertain and values in the range 3 to 5 kpc
have been proposed.

Both radio and X-ray emission coincident with this GC has been also reported by different authors \cite{kob05, p07}.
The marginal and extended radio detection comes from the NVSS survey with an integrated flux density of $20.5\pm 3.6$ mJy
at 20 cm. Inspection of the Very Large Array (VLA) archive reveals data sets at the GC position obtained in 1990 at the same wavelength
but using the B array configuration, which provides better angular resolution than the NVSS.
We recalibrated them in order to
produce a radio map with high angular resolution. As a result no compact radio sources were detected above four
times the rms noise of 0.25 mJy beam$^{-1}$. This fact suggest that the radio emission is intrinsically extended or
resulting from the combined effect of faint point-like radio sources.

The X-ray emission observed with the {\it Chandra} satellite \cite{hei05, p07} is well resolved into many point-like sources
inside the GC radius together with a diffuse component. These objects are most likely a mixture of
cataclysmic variables, quiescent Low-mass X-ray binaries (LMXB) and MSPs, among other objects.
The intrinsic total X-ray luminosity of the GC in the 0.5-8 keV band
is estimated to be $\sim 2 \times 10^{33}$ erg s$^{-1}$.

The finding of a GC consistent with a bright Fermi source
is a remarkable fact that deserves a careful attention. 
Beyond such positional coincidence, the key issue in order to claim
a possible association is the availability or not of a physical scenario consistent with
the observed gamma-ray flux. As quoted in Section 1, expectations of the gamma-ray emission from GCs
are available in the literature
\cite{bs07}. The key model parameters are the spectral index of the power law energy distribution for the
leptons injected by the MSP population ($\alpha$), the GC stellar luminosity ($L$), the lepton energy cutoffs,
the energy conversion efficiency
($\eta \simeq 0.01$), the pulsar surface magnetic field (usually $B=10^9$ G) and spin period (usually a few ms).
The magnetic field inside the GC is fixed to $10^{-6}$ G and their adopted number of MSPs is $N_p \simeq 100$.

In Fig. \ref{comparison}, we plot the theoretical predictions
together with the observed spectrum for the two Fermi sources: \object{0FGL J1848.6$-$0138} in discussion here and
the similar \object{0FGL J0025.1$-$7202}. The latter is likely related to the GC 47 Tuc
specifically modelled by Bednarek \& Sitarek (2007).
Given that it seems reasonable to initially assume that a similar
emission mechanism could be at work in both clusters GLIMPSE-C01 and 47 Tuc,
we have scaled the same model to their conceivable distances of 3 and 4 kpc.
The \object{0FGL J1848.6$-$0138} spectrum can be represented by
$N($ph~erg$^{-1}$~cm$^{-2}$~s$^{-1})= 2.40 \times 10^{-8}[E/{\rm GeV}]^{-2.14}$.
This is simply the result of fitting  a simple power law spectrum to the Fermi
gamma-ray flux measurements in the 0.1-1 GeV and 1-100 GeV bands \cite{fermicat}.
The lepton energy limits are between $1$ and $3 \times 10^4$ GeV. A similar procedure
has been followed for  \object{0FGL J0025.1$-$7202}.
Based on the available Fermi fluxes, it seems that their parameter set
with $\alpha=2$, $L=7.5 \times 10^5$ $L_{\odot}$ and a low energy
cutoff $E_{\rm min} = 1$ GeV provides the closest theoretical
prediction, although both Fermi spectra
appear to significantly exceed the model.

The non-perfect agreement in this qualitative comparison can be due to several different effects
not correctly taken into account. For instance, the contribution to the gamma-ray spectrum at low energies from
scattering of the microwave background radiation could not be negligible in the case of GLIMPSE-C01, whose stellar luminosity
($L\simeq 10^5$ $L_{\odot}$) is not as high as in the 47 Tuc case. In addition, we cannot completely
exclude that the distance to GLIMPSE-C01 has
been overestimated since this key parameter is very difficult to determine in a heavily absorbed case such as this.
Despite these problems, the possibility for GLIMPSE-C01 being a Fermi gamma-ray source
appears as a plausible one when considering all the parameter uncertainties we have just mentioned.

In order to provide a distance independent indicator of the emission mechanism, it is instructive to compare
the X-ray source counts in the GLIMPSE-C01 and 47 Tuc case. Indeed, the cluster population
of X-ray binaries are believed to be the direct progenitors of the gamma-ray 
emitting MSPs (see e.g. Bhattacharya (1996)
for a review).
Pooley et al. (2007) report 13 sources with unabsorbed 0.5-8 keV X-ray luminosity above $10^{31}$ erg s$^{-1}$.
In contrast, the comprehensive X-ray survey of 47 Tuc by Heinke et al. (2005)
yielded nearly 3 times more sources
above a similar luminosity and energy range. Thus, despite Pooley et al. (2007)
infer a high production rate
of X-ray binary systems through close stellar encounters, this is not observationally translated into
a significantly enhanced X-ray source population.

\begin{table*}
\caption{X-ray sources with point-like infrared counterparts inside the 0FGL J1848.6$-$0138 error circle}
\label{xray}
\begin{center}
\begin{tabular}{ccccccccc}
\hline
\hline
2XMM              &  Energy flux            &  X-ray/IR  &  $J$  &  $H$  & $Ks$  &  $3.6 \mu$m &  $4.5 \mu$m & $ 5.8 \mu$m  \\
source name       &  (0.5-4.5 keV)          &  offset          &       &       &       &             &             &                         \\
                  & $10^{-15}$ erg s$^{-1}$ cm$^{-2}$  &  $^{\prime\prime}$        & mag   &  mag  & mag   &  mag        &   mag       &    mag         \\
\hline
J184852.3-014026 &  $26 \pm 4$    & 2.4 & $ 7.77 \pm 0.02$ & $ 7.21 \pm 0.05$& $ 6.97 \pm 0.03$ & $ 6.92 \pm 0.04$ & $ 6.95 \pm 0.04$ & $ 6.89 \pm 0.03$   \\
J184813.2-014427 &  $7.6 \pm 2.9$ & 1.8 & $\geq 16.61$     & $14.75 \pm 0.08$& $12.99 \pm 0.04$ & $11.76 \pm 0.05$ & $11.49 \pm 0.07$ & $11.22 \pm 0.11$   \\
J184805.0-013726 &  $5.8 \pm 1.4$ & 1.8 & $\geq 16.65$     & $\geq15.19     $& $13.08 \pm 0.05$ & $10.59 \pm 0.06$ & $ 9.76 \pm 0.07$ & $ 9.28 \pm 0.05$   \\
\hline
\end{tabular}
\end{center}
\end{table*}

Given the evolutionary connection between X-ray binaries and MSPs, the cluster X-ray luminosity is
believed to roughly scale to the total number of MSPs. We have therefore computed the cluster
X-ray to gamma-ray luminosity ratio according to $L_{0.3-8~{\rm kev}}/L_{0.1-1~{\rm GeV}}$ based on
the observational data quoted above. The resulting value is $\sim 10^{-4}$ for 47 Tuc and $\sim 10^{-5}$ for GLIMPSE-C01.
The fact that this ratio is smaller by at least an order of magnitude in GLIMPSE-C01
would seem to go against its identification with the Fermi source.
The total number of MSP in 47 Tuc is estimated to be $\sim50$ \citep{bog06,fermidetection}.
Thus scaling with the X-ray source luminosity one would expect an smaller value of $\sim20$ in the GLIMPSE C01 case.
Nevertheless, we cannot strictly rule out a similar gamma-ray production mechanism in both clusters
that provides a clear gamma-ray detection with different luminosities in future more sensitive observations.

Alternative scenarios to the one discussed above for GC gamma-ray emission
can also be considered. In particular,
we cannot exclude that other emission mechanisms
are at work inside the GC such as an intermediate massive black hole in its centre, peculiar LMXBs, etc.
Gamma-ray variability would be likely expected in this context, but no evidence of it has been obtained until now.

\subsection{A possible PWN as a counterpart?}

We have also explored the possibility that the Fermi source 
is associated to any other peculiar object inside its 95\% confidence radius.
One of them, uncatalogued in the SIMBAD database, is almost at the centre
of the Fermi error box with an apparent bubble-like shape already mentioned.
Its angular diameter extends  
about $2^{\prime}$ as illustrated in the GLIMPSE image of Fig. \ref{cluster_image}.

This object is also very well detected in the radio NVSS images with
a 20 cm integrated flux density of $88 \pm 4$ mJy and its morphology is reminiscent of a PWN. 
Radio emission from this bubble feature is shown in detail in the Fig. \ref{cluster_image} right panel
but no X-ray detection is obtained when inspecting XMM archival data. The resulting
X-ray flux upper limit (3-$\sigma$) in the 0.5-4.5 keV band is estimated as 
$6\times 10^{-14}$ erg s$^{-1}$ cm$^{-2}$ for the region covered by the putative PWN. The lack
of X-ray detection is difficult to reconcile with a PWN interpretation unless we are dealing with an old, evolved
pulsar that has already deposited all its spindown power into the nebula \cite{jag09}.

As an alternative
possibility, a newly discovered bubble blown by a central star could be
considered as well.
The stellar-like object closest to the shell centre that we would propose as the most likely exciting
source of the shell-like structure
is located at $RA=18^h 48^m $43\rl 72 and $DEC=-01^{\circ} 38^{\prime}$38\pri 1 with
$Ks=13.21$ mag. Its colours in the 2 Micron All Sky Survey (2MASS)
are suggestive of a very reddened star  ($J-Ks \simeq +4.3$).
In such a case, we speculate on a possible hadronic interaction in the
shocked region of the gas shell that would require further attention.

\subsection{An ultracompact HII region in the field}

Another remarkable object inside the Fermi error circle is the
bright radio source \object{GPSR5 31.243$-$0.110} likely to be an ultracompact HII region  \citep{giv07}
based on its morphology. Its gamma-ray emitting nature is not clear given the lack of suitable physical scenarios
for this kind of objects.

\subsection{X-ray emitting stellar-like objects in the field}

Several stellar-like objects with X-ray counterparts are also present inside
the Fermi error circle as evidenced by the comparison of GLIMPSE and XMM catalogue shown 
in Fig. \ref{cluster_image}. None of them is an NVSS radio source.
Their observational properties are listed in Table \ref{xray}.
We cannot rule out that any of these stellar-like objects is behind the gamma-ray
source taking into account that a significant fraction of Fermi sources in the
galactic plane could be related to pulsars both isolated and in binary systems.

\section{Conclusions}

We have reported an extensive
search for counterparts of the unassociated source \object{0FGL J1848.6$-$0138}.
As a result, we find that this the second Fermi gamma-ray source
with a possible association with a GC.
The emission level observed by Fermi is not perfectly explained by previous theoretical
models based on leptons accelerated by the MSP population inside a GC and comptonizing the
stellar and microwave background radiation. However, the disagreement between
current theories and observation is within an order of magnitude and this fact does not rule out that
a consistent physical scenario is conceivable by means of this physical mechanism. Improved theoretical models and
better estimates of the cluster physical parameters (specially the distance) will be required to
resolve such apparent discrepancies and, perhaps, confirm the idea of GCs as gamma-ray sources.

In addition to the GC scenario, several other peculiar objects inside the Fermi
error circle stand for alternative counterpart candidates.
The most interesting of them is very close to the circle centre and resembles a PWN in infrared and radio images.
However, the lack of obvious X-ray emission makes its true nature not so clear.
Alternatively, it could also be a more ordinary stellar, wind-blown bubble.

Future Fermi observations will certainly narrow the position uncertainty of the gamma-ray
source thus enabling us to exclude or confirm some of the counterpart candidates reported here.

\begin{acknowledgements}
The authors acknowledge support by
grant AYA2007-68034-C03-02
from the Spanish government, and FEDER funds.
This has been also supported by Plan Andaluz de Investigaci\'on
of Junta de Andaluc\'{\i}a as research group FQM322.
J.A.C. is a research member of the Consejo Nacional de Investigaciones
Cient\'{i}ficas y Tecnol\'ogicas (CONICET), Argentina.
The NRAO is a facility of the NSF
operated under cooperative agreement by Associated Universities, Inc.
This research made use of the SIMBAD
database, operated at the CDS, Strasbourg, France.
This publication makes use of data products from the Two Micron All Sky Survey, which is a
joint project of the University of Massachusetts and the Infrared
Processing and Analysis Center/California Institute of Technology,
funded by the National Aeronautics and Space Administration and the
National Science Foundation in the USA.
We also thank an anonymous referee for helping us to significantly improve this paper.
\end{acknowledgements}

\bibliographystyle{aa} 
\bibliography{Yourfile} 

\begin{thebibliography}{}

\bibitem[Abdo et al. 2009a]{fermicat} Abdo, A. A., et al. 2009a, \apjs, 183, 46
\bibitem[Abdo et al. 2009b]{fermidetection} Abdo, A. A., et al. 2009b, Science, 325, 845
\bibitem[Atwood et al. 2009]{at09} Atwood, W. B., et al. 2009, \apj\ (submitted) arXiv:0902.1089 {\tt astro-ph}
\bibitem[Bednarek \& Sitarek  2007]{bs07} Bednarek, W. \& Sitarek, J. 2007, \mnras, 377, 920
\bibitem[Benjamin et al. 2003]{b03} Benjamin, R. A., et al. 2003, \pasp, 115, 953
\bibitem[Bhattacharya 1996]{bh96} Bhattacharya, D., 1996, ASP Conference Series, 105, 547
\bibitem[Bogdanov et al. 2006]{bog06} Bogdanov, S. et al. 2006, \apj, 646, 1104
\bibitem[Chen 1991]{chen91} Chen, K. 1991, \nat, 352, 695
\bibitem[Condon et al. 1998]{con98} Condon, J. J., et al. 1998, \aj, 115, 1693
\bibitem[de Jager et al. 2009]{jag09} de Jager, O. C., et al. 2009, Proc. of the 31$^{th}$ ICRC, L\'od\'z
arXiv:0906.2644v1  {\tt astro-ph}
\bibitem[Giveon et al. 2007]{giv07} Giveon, U. et al. 2007, \aj, 133, 639
\bibitem[Heinke et al. 2005]{hei05} Heinke, C. O., et al. 2005, \apj, 625, 796
\bibitem[Kobulnicky et al. 2005]{kob05} Kobulnicky, H. A., et al. 2005, \aj, 129, 239
\bibitem[Pooley et al. 2007]{p07} Pooley, D., et al. 2007, arXiv:0708.3365v1 {\tt astro-ph}
\bibitem[Tavani 1993]{tav93} Tavani, M., 1993, \apj, 407, 135

\end{thebibliography}

\end{document}